\documentclass[journal]{IEEEtran}

\usepackage{subfigure}
\usepackage{graphicx}
\usepackage{latexsym}
\usepackage[fleqn]{amsmath}
\usepackage{amsmath}
\usepackage{CJK}
\usepackage{indentfirst}
\usepackage[varg]{txfonts}
\usepackage{stfloats}
\usepackage{multirow}%
\usepackage{booktabs}
\usepackage{color}
\usepackage{color}
\usepackage{epstopdf}
\usepackage{float}
\usepackage{bm}
\usepackage[linesnumbered,ruled,commentsnumbered,longend]{algorithm2e}
\makeatletter

\newcommand{\Rmnum}[1]{\expandafter\@slowromancap\romannumeral #1@}
\makeatother

\ifCLASSINFOpdf
\else
\fi
\begin{document}
%
\title{{Ultra Wide Band THz IRS Communications: Applications, Challenges, Key Techniques, and Research Opportunities}}
\author{Wanming Hao,~\IEEEmembership{Member,~IEEE,} Fuhui Zhou,~\IEEEmembership{Senior Member,~IEEE,} Ming Zeng,~\IEEEmembership{Member,~IEEE,} Octavia A. Dobre,~\IEEEmembership{Fellow,~IEEE,} and~Naofal Al-Dhahir~\IEEEmembership{Fellow,~IEEE}
\thanks{W. Hao is with the School of Information Engineering, Zhengzhou University, Zhengzhou 450001, China. (iewmhao@zzu.edu.cn)}
\thanks{F. Zhou is with the College of Electronic and Information Engineering, Nanjing University of Aeronautics and Astronautics,
	Nanjing, 210000, China. (email: zhoufuhui@ieee.org)}
\thanks{M. Zeng is with the Department of Electrical Engineering and Computer Engineering, Université Laval, Quebec, G1V 0A6, Canada. (E-mail: ming.zeng@gel.ulaval.ca)}
\thanks{O. A. Dobre is with the Faculty of Engineering and Applied Science, Memorial University, St. Johns, NL A1B 3X5, Canada. (E-mail:  odobre@mun.ca)}
\thanks{N. Al-Dhahir is with the Department of Electrical and Computer Engineering, University of Texas, Dallas, USA. (E-mail:  aldhahir@utdallas.edu)}}
\maketitle
\begin{abstract}
Terahertz (THz) communication is a promising technology for future wireless networks due to its ultra-wide bandwidth. However, THz signals suffer from severe attenuation and poor diffraction capability, making it vulnerable to blocking obstacles.  To compensate for these two shortcomings  and improve the system performance, an intelligent reflecting surface (IRS) can be exploited to change the propagation direction and enhance the signal strength. In this article, we investigate this promising ultra wide band (UWB) THz IRS communication paradigm. We start by motivating our research and describing several potential application scenarios.  Then, we identify major challenges faced by UWB THz IRS communications.   To overcome these challenges, several effective key techniques are developed, i.e., the time delayer-based sparse radio frequency antenna structure, delay hybrid precoding and IRS deployment. Simulation results are also presented to compare the system performance for these proposed techniques, thus demonstrating their effectiveness. Finally, we highlight several open issues and research opportunities for UWB THz IRS communications.
\end{abstract}


%
\IEEEpeerreviewmaketitle

\section{Introduction}
High transmission speed  and low power consumption are two key performance metrics for future wireless communications. For example, more than 1 Tbps peak data rate should be supported in the ultra-high-definition video transmissions and more than 10 fold increase in  energy efficiency should be obtained for the internet of things (IoTs)~\cite{Ding_2019_VTM}.  To satisfy the above requirements, terahertz (THz, 0.1$\sim$10 THz)  and intelligent reflecting surface (IRS) are two promising candidate technologies. THz communication can provide ultra-wide bandwidth to realize the extremely-high transmission rate~\cite{Dai_2021_JSAC}. However, the coverage of THz signals is limited due to the severe path loss and poor diffraction ability. To this end, an IRS consisting of a large number of reconfigurable passive elements can be applied to change the propagation direction and enhance the signal strength by adjusting the phase/amplitude of the IRS elements, thereby improving the coverage of the THz signals~\cite{Zhang_2019_TWC}.  This motivates us   to investigate the ultra wide band (UWB) THz IRS communications. 

Because of the small wavelength of the THz signals, the base station (BS) can be equipped with a large number of antennas within limited physical space  to improve the beamforming gain. Meanwhile, to reduce the radio frequency (RF) power consumption, a sparse RF chain antenna structure is usually applied, where a few RF chains are connected to all antennas by low-power consumption phase shifters (PSs)~\cite{Yan_2020_JSAC}. Furthermore, for the sparse RF chain-based fully-connected and static/dynamic subarray structures, current research results have shown that  near optimal performance can be obtained by optimizing the analog/digital precoding. However, unlike the low frequency or narrow band system, the beam directions at different subcarriers  may vary for the UWB THz system, which is  called beam split~\cite{Dai_2019_BLOBECOM}.  Due to the frequency independent characteristic of the PSs for the sparse RF chain antenna structure, there exists severe loss in beam gain when the conventional hybrid precoding schemes are applied. IRS further leads to   two stage beam split for UWB THz IRS communications, namely, the first stage beam split from the BS to the IRS and the second stage  beam split from the IRS to the receivers. Such beam split effects are also affected by the actual IRS deployments.  As a result, the conventional IRS deployment and BS antenna structures are not suitable for UWB THz IRS communications, calling for novel solutions and structures.

In this article, we investigate the following aspects of UWB THz IRS communications. First, we motivate our research and introduce three potential application scenarios, namely, indoor communications, dense building communications, and unmanned aerial vehicle (UAV) communications. Next, we summarize the state-of-the-art  on the UWB THz IRS communications. Then, we analyze the major challenges underlying UWB THz IRS communications. To deal with these challenges, we propose several key techniques, including the time delayer (TD)-based sparse RF antenna structure, delay hybrid precoding, and IRS deployment. Furthermore, we provide  simulation results to compare the system performance under different techniques. Finally, we highlight several open problems and potential research opportunities for UWB THz IRS communications, such as artificial intelligence (AI)/machine learning (ML)-empowered techniques, selection of the TDs, semantic communications and sensing and localization. Our main goal is to provide useful technical guidance and new research topics for future 6G  wireless communications. 

The rest of this article is organized as follows: The research motivations, applications and state-of-the-art are presented in Section II. The major challenges for UWB THz IRS communications are summarized in Section III, whereas the corresponding key techniques are given in Section IV. Performance results are presented in Section V. Finally, the open issues and research opportunities are highlighted in Section VI, while conclusion is drawn this article in Section VII. 
\section{Motivations, Applications, and State-of-the-Art}
\begin{figure*}[t]
	\begin{center}
		\includegraphics[width=14cm,height=7cm]{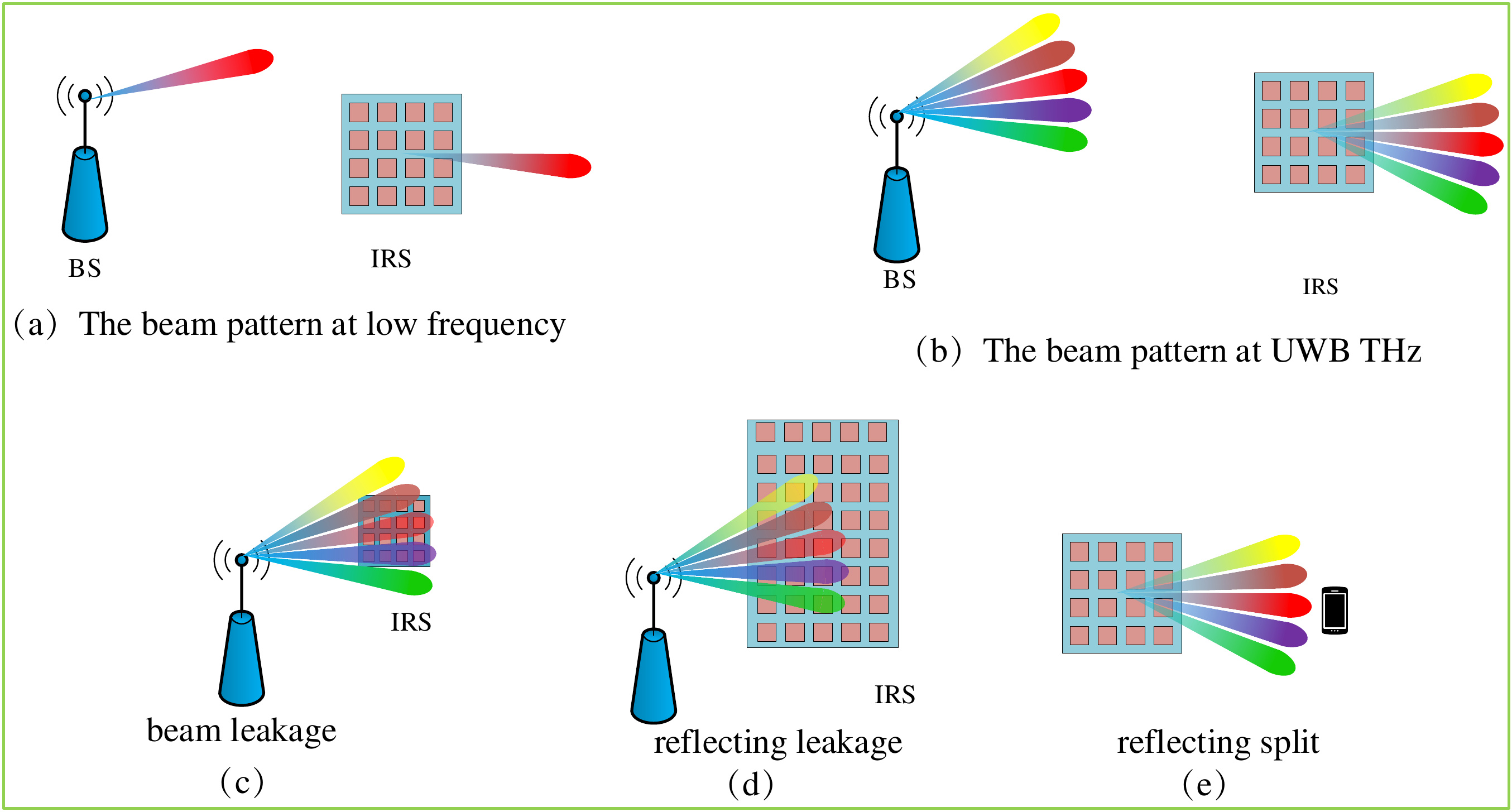}
		\caption{The beam patterns and the beam split effects.}
		\label{figure1}
	\end{center}
\end{figure*}

In this section, we first provide the research motivations, and then introduce several potential application scenarios. Finally, we summarize the state-of-the-art for THz IRS communications.   
\subsection{Motivations}
Recently, THz communication has attracted considerable attention owing to its ability to provide ultra-wide bandwidth. However, due to its extremely high frequency, THz signals usually suffer from severe attenuation and poor diffraction, which limits its transmission distance and applications in mobile communications. Although the multiple-input multiple-output (MIMO) technique can be applied to design high gain directional beamforming to improve the signals strength, THz signals are still vulnerable to  blocking obstacles, which seriously  degrades the beamforming gain when obstacles exist. Therefore, it is essential for THz communications to have line-of-sight (LoS) links between the transmitter and receiver. Motivated by this consideration, IRS can be deployed to create the additional LoS links by judiciously reflecting the received signals, and thus, enhancing the THz signals strength and extending the coverage of the THz communications. 
\subsection{Application Scenarios}
Based on the above discussions, next we briefly introduce three potential application scenarios for UWB THz IRS communications.
\subsubsection{Indoor Communications}
UWB THz is an attractive candidate frequency band to provide the ultra-high transmission rate for the indoor short distance communications. Indeed, the 0.252-0.325 THz has been selected by the IEEE 802.15.3d standard to support tens of Gbps data rate~\cite{IEEEStandard}. However, the existence of various obstacles in indoor environments often makes it difficult to build a LoS link. The deployment of IRS, e.g. on an indoor wall or ceiling, can help create a virtual direct LoS link, and thus, improving the quality-of-service of users.  
\subsubsection{Dense Building Communications} 
As mentioned earlier, THz signals are vulnerable to blocking obstacles. Thus,  the presence of several dense buildings makes it difficult to build a direct link, or such a link will be very weak. Therefore, it is necessary to apply IRS to change the signals transmission direction so as to establish a virtual direct link between the transmitter and receiver, improving the strength of the received signals.  
\subsubsection{UAV Communications}
Recently, UAV communications have received much attention from both academia and industry due to their high flexibility and low deployment cost. Furthermore, LoS links between the UAV and ground users can usually be built to  improve the system performance. However, conventional UAV are typically equipped with active antennas, and the power consumption can be very huge, especially with a large number of antennas. This often results in a limited service time due to the  battery-powered UAV.  By integrating the UAV with an energy-efficient IRS, forming a THz UAV IRS, one does not only provide ultra high rate but also decreases the power consumption and prolongs the service time. In UAV-based systems, the IRS can be deployed at the UAV or on the ground. 
\subsection{State-of-the-Art}
{There are only few works considering THz IRS communications~\cite{Ma_2020_Access}-\cite{Qiao_2020_WCL}. Towards decreasing the power consumption, the authors in~\cite{Ma_2020_Access} apply the sparse RF
chain antennas structure at the BS. Furthermore, a joint channel estimation and hybrid precoding scheme is proposed to maximize the system throughput. Similarly, to obtain the channel state information (CSI), an effective  beam training scheme is developed in~\cite{Ning_2021_TVT}. Consequently, the corresponding hybrid beamforming is designed for maximizing the transmission rate. UAV-assisted THz IRS communications is investigated in~\cite{Pan_2021_WCL}, where the authors consider a single-antenna BS to analyze and optimize the UAV trajectory. The authors in~\cite{Qiao_2020_WCL} study the secure transmission problem, and jointly design the BS active beamforming and IRS passive beamforming to improve the system security.}

 \begin{figure*}[t]
 	\begin{center}
 		\includegraphics[width=16cm,height=4.5cm]{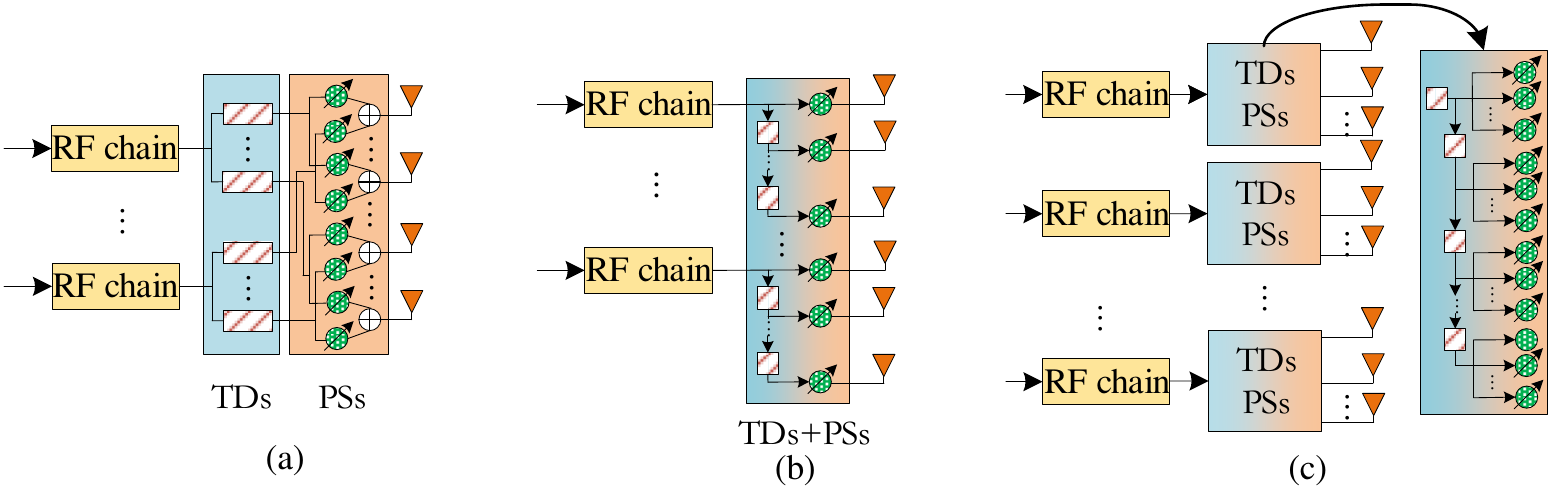}
 		\caption{Three TD-based sparse RF chain antenna structures.}
 		\label{figure2}
 	\end{center}
 \end{figure*}
 
{Current works mainly focus on the narrow band systems, where the research methods or adopted techniques are similar to those proposed in conventional low frequency or millimeter wave IRS communications. Since the THz band enjoys  the ultra-wide bandwidth, it is important to study new challenges and techniques for the UWB THz IRS communications. In particular,  we will analyze whether the conventional sparse RF antenna structures are still suitable. Major challenges and key techniques are also introduced.}  
\section{Major Challenges}
In this section, we identify the major challenges faced by UWB THz IRS communications, including the huge power consumption and the high loss in  beam gain.
\subsection{Huge Power Consumption for the Active Antennas Elements at the BS} 
Generally, for THz systems, the BS is equipped with a large number of antennas to form the high gain directional beam for improving the strength of the THz signals. However, when the BS is equipped with a large scale antenna array, the hardware power consumption is huge for the THz systems.  This is because the
power consumption for a single RF chain at THz frequency (about 250 mW) is almost ten times higher than that at low frequency (about 30 mW) ~\cite{Heath_2016_JSTSP}. If each antenna is connected to a unique RF chain as in the traditional structure, the large number of  antennas will lead to a huge power consumption. Therefore, several sparse RF chain antenna structures have been developed, where the number of RF chains is much lower than that of the antennas. In this case,  multiple antennas are  connected to one RF chain via several low-power consumption PSs, such as the fully-connected structure and static/dynamic subarray structure. Although the existing sparse RF chain antenna structures are very effective in decreasing the power consumption for the mmWave or narrowband systems without causing much system performance loss, they may not be as effective for the  UWB THz IRS systems. The reasons are as follows: the beam directions at different subcarriers  are often different (i.e., beam split), and thus, yielding severe loss in beam gain under the conventional hybrid precoding schemes due to the frequency independent characteristic of the PSs. Therefore, how to obtain a high system performance with minimum power consumption is a major challenge.
\subsection{The Large Loss in Beam Gain Due to the Two Stage Beam Split} 
{Different from the low frequency narrowband systems, there exists beam split for the UWB THz systems, namely, beam directions produced by the BS at different subcarriers are different~\cite{Dai_2019_BLOBECOM}. Furthermore, as shown in Fig.~\ref{figure1}, the UWB THz IRS systems face a two stage beam split, including the first stage beam split from the BS to the IRS  and the second stage beam split from the IRS to users.}

Generally, the first stage beam split may cause two problems as shown in Figs.~\ref{figure1}(c) and 1(d), namely  ``beam leakage'' and ``reflecting leakage''. The former means that the beam directions formed at several subcarriers deviate from the IRS, so that these THz signals cannot be reflected by the IRS, leading to a loss in the beam gain. The latter means that some of the IRS elements do not reflect any THz signals due to the small beam split, which decreases the reflecting efficiency. Note that the ``beam leakage'' or ``reflecting leakage'' is related to several factors, such as the number  of the BS antennas, the IRS shape, the distance between the BS and IRS, and the carrier frequency.  In addition, since the size of the receivers is relatively small, the second stage beam split leads to a loss in the beam gain, as shown in Fig.~\ref{figure1}(e), namely, the beam directions reflected by the IRS at several subcarriers deviate from the receivers, and we call this ``reflecting split''. Clearly, the two stage beam split may cause high loss in array gain  for UWB THz IRS communications, and how to overcome the two stage beam split effects is another major challenge.

\section{Key Techniques}
To overcome the first stage beam split, we  introduce three TD-based sparse RF chain antenna structures and the corresponding hybrid precoding that can realize the beam convergence and dispersal. Following this, we analyze the effect of the IRS deployment on the beam split.      

\subsection{TD-Based BS Antenna Structure}
As mentioned earlier, for UWB THz IRS communications, how to overcome the two stage beam split effects is a major challenge, and the existing sparse RF chain antenna structures are not effective.  Recently, it was shown that the TDs can be used to mitigate the beam split effect, and thus, several TD-based sparse RF chain antenna structures have been developed~\cite{Dai_2019_BLOBECOM,Zhai_2021_JSAC}, as shown in Fig.~\ref{figure2}. First, the authors in \cite{Dai_2019_BLOBECOM} propose a fully-connected structure, where the PSs are connected to the RF chain via several TDs~(i.e., Fig.~\ref{figure2}(a)). This structure can obtain a better performance in restraining the beam split, but the hardware complexity is high and the number of the PSs is also huge. Later, the authors in~\cite{Zhai_2021_JSAC} propose a subarray structure, where each PS (or antenna) is connected to only one TD~(i.e., Fig.~\ref{figure2}(b)). Although the hardware complexity of the subarray structure is low, it needs more TDs. Note that the power consumption for the THz TDs (e.g., 80 mW~\cite{Cao_2016_IEEEJQE}) is much higher than that for the PSs (e.g., 30 mW). Under a large number of antennas, the number of the TDs is also large, which leads to the huge power consumption.  {To address this issue, we develop a sparse TD-based subarray structure as shown in Fig.~\ref{figure2}(c), where multiple PSs (or antennas) are connected to one TD. In this way, the number of TDs can be effectively reduced, together with the power consumption and hardware complexity. In addition, compared with the first structure, all TDs can set the same delay for our designed structure, thereby simplifying the TD circuit control and design.}
\subsection{Delay-Based Hybrid Precoding}
For the TD-based sparse RF chain antenna structure, the analog precoding, delay and digital precoding should be jointly optimized. In fact, the main objective of the TDs is to overcome the beam split effect.  \cite{Dai_2019_BLOBECOM} first introduces how to realize the beam convergence by adjusting the TDs, while~\cite{Zhai_2021_JSAC} studies how to disperse the beam by adjusting the TDs for serving more users. Different from the existing work, two beam split effects (``beam leakage'' and ``reflecting leakage'') need to be solved in the UWB THz IRS communications. In this way, we should first design the analog beamforming at the BS by adjusting the PSs to point the beam direction of the central subcarrier towards the IRS. For the ``beam leakage'' case,  the overall beams should be converged by adjusting the TDs so that the IRS can reflect all received beams. For the ``reflecting leakage'' case, the overall beams should be dispersed by adjusting the TDs so that the beams can  cover the entire IRS, thus improving the reflecting efficiency. Finally, the digital precoding should be designed to effectively integrating multi-user interference. In addition,  beam split effects depend on the distance between the BS and IRS, the number of the BS antennas, the IRS deployment, the carrier frequency, etc.  With the above procedure, the beam split challenge can be effectively mitigated using the delay-based hybrid precoding. 

\subsection{IRS Deployment}
\begin{figure}[t]
	\begin{center}
		\includegraphics[width=7cm,height=7cm]{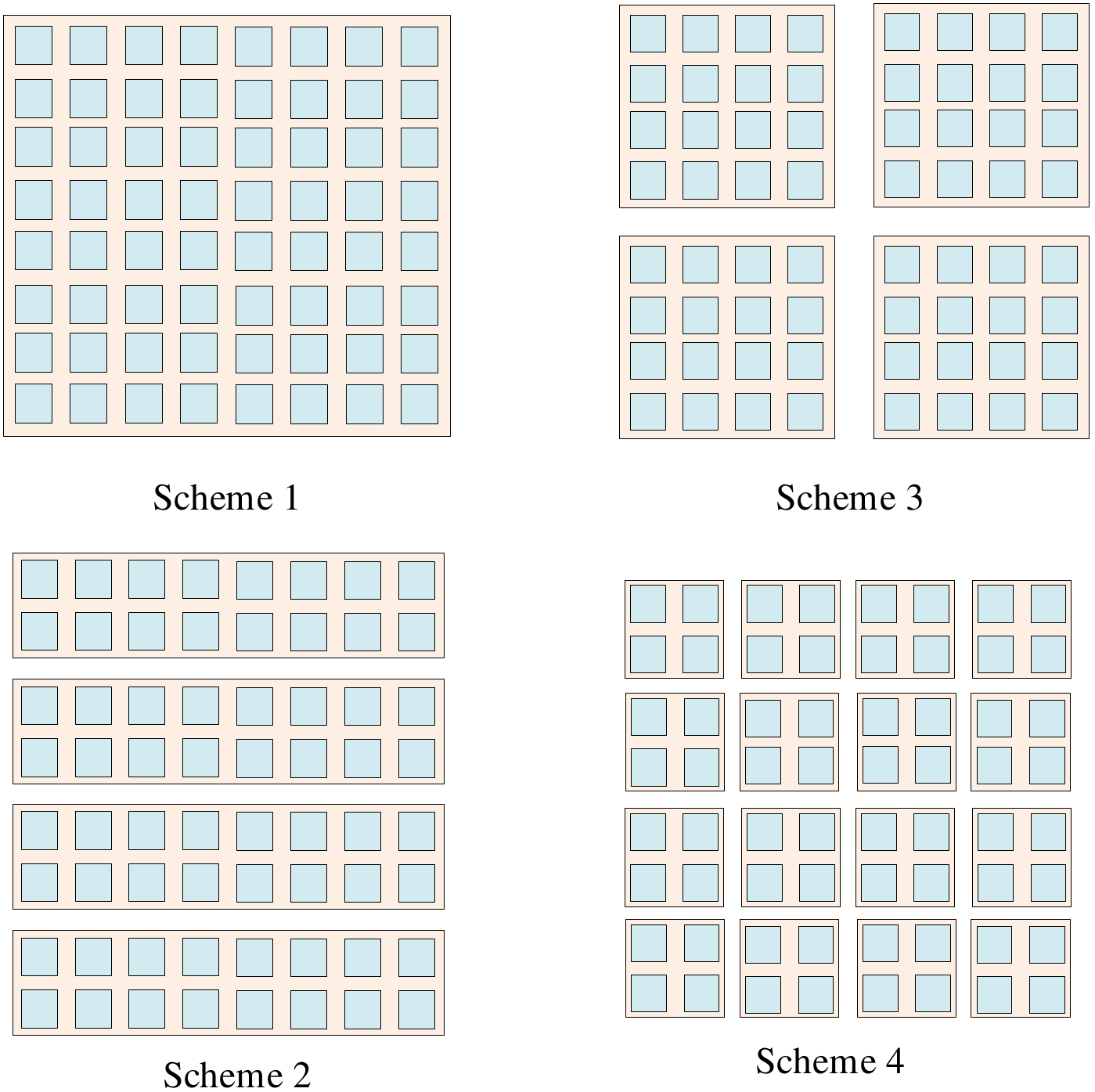}
		\caption{Different IRS deployment schemes.}
		\label{figure3}
	\end{center}
\end{figure}
Before discussing the IRS deployment, we first analyze the reason for the beam split effects in UWB THz signals. With a large number of antennas at the transmitter or receiver, not all antennas of the receiver simultaneously receive the signals, namely, there exists signal delays among different antennas. In addition, for the low frequency multiple carriers, the ratio $f_m/f_c$ can be approximated as 1, where $f_m$ and $f_c$ stand for the $m$th and central subcarrier frequency, respectively. However,  for the UWB THz, the above approximation will not hold. Meanwhile, the selection of the space among adjacent antennas  usually depends on the wavelength of the central subcarrier frequency. Based on the above reasons, the beamforming of the different subcarrier signals will generate different directions referred to as beam split, and the detailed analysis can be found in~\cite{Gao_2021_JSAC}.  Furthermore, more antennas (elements) at the BS (IRS) lead to a larger signal delay among different antennas (elements) at the BS (IRS), which further amplifies the beam splits effects.  Generally, the IRS is a planar structure,  and the horizon and elevation directions can both form the beam split, aggravating the effects. Thus, the IRS deployments (sizes) also contribute to the beam split effects. As shown in Fig.~\ref{figure3}, the large central IRS deployment (i.e., Scheme 1), distributed multiple rectangular IRS deployment (i.e., Scheme 2), distributed multiple square IRS (i.e., Scheme 3) and  distributed large scale square IRSs deployment  (i.e., Scheme 4),  lead to different beam split effects. Our research finds that Scheme 4 owns the smallest beam split effects, but the deployment cost will be high.  Detailed performance analysis is presented below. 
\section{Performance Analysis}
\begin{figure}[t]
	\begin{center}
		\includegraphics[width=8.4cm,height=4.5cm]{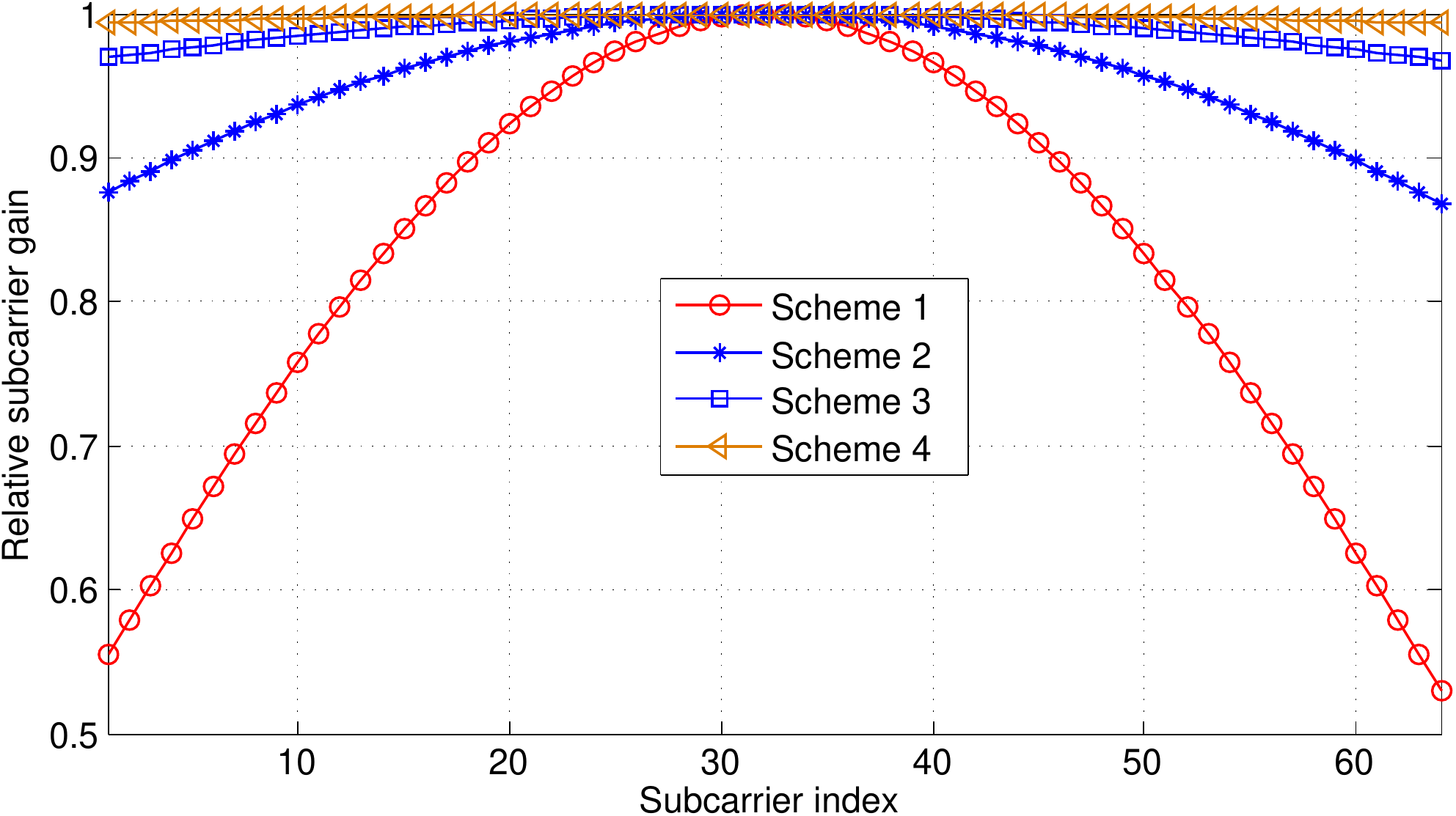}
		\caption{Relative subcarrier gain under different IRS deployment schemes.}
		\label{figure4}
	\end{center}
\end{figure}
\begin{figure}
	\centering
	\subfigure[]{
		\label{figure51}	
		\includegraphics[width=8.7cm,height=3.6cm]{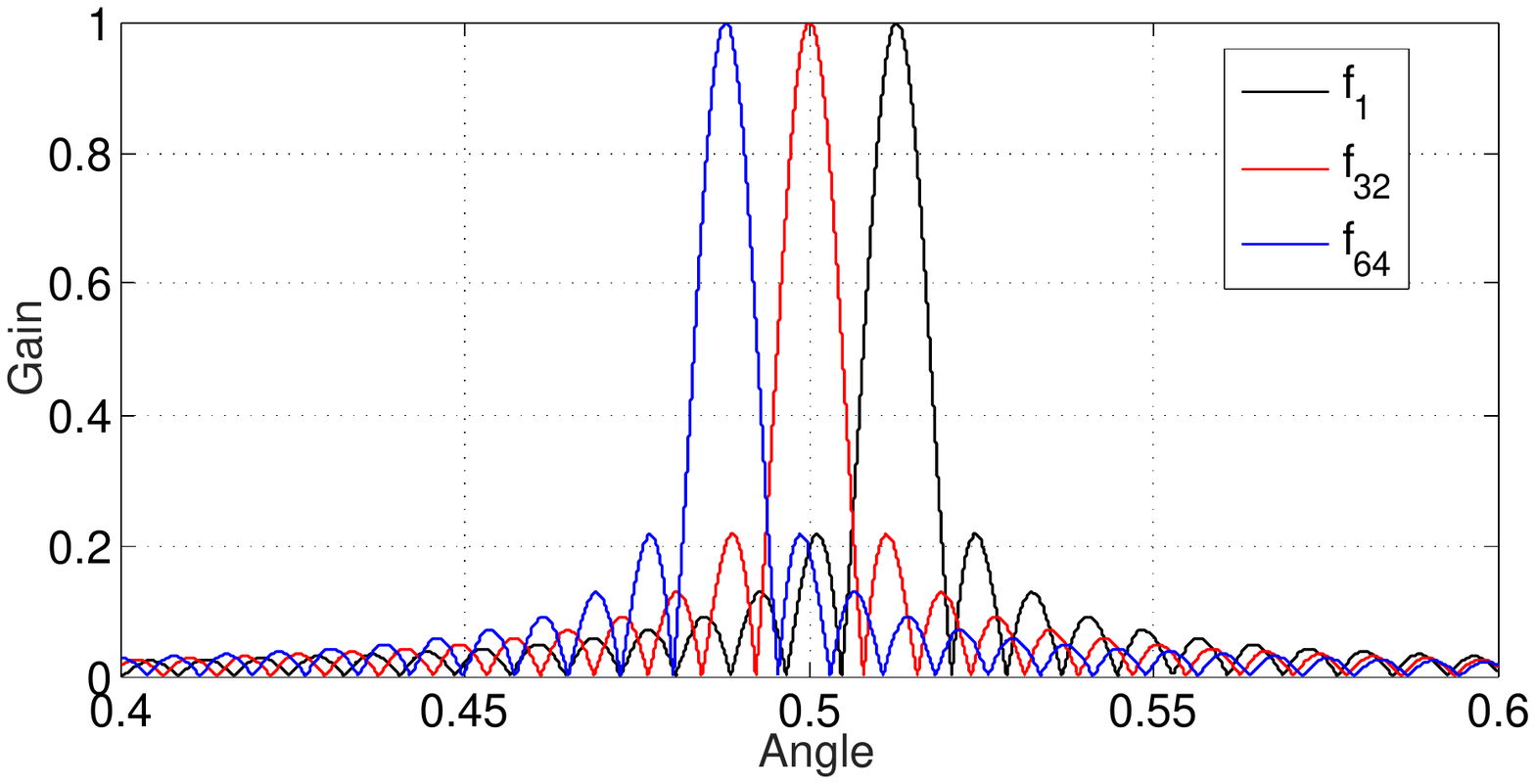}}
	\subfigure[]{
		\label{figure52}	
		\includegraphics[width=8.7cm,height=3.6cm]{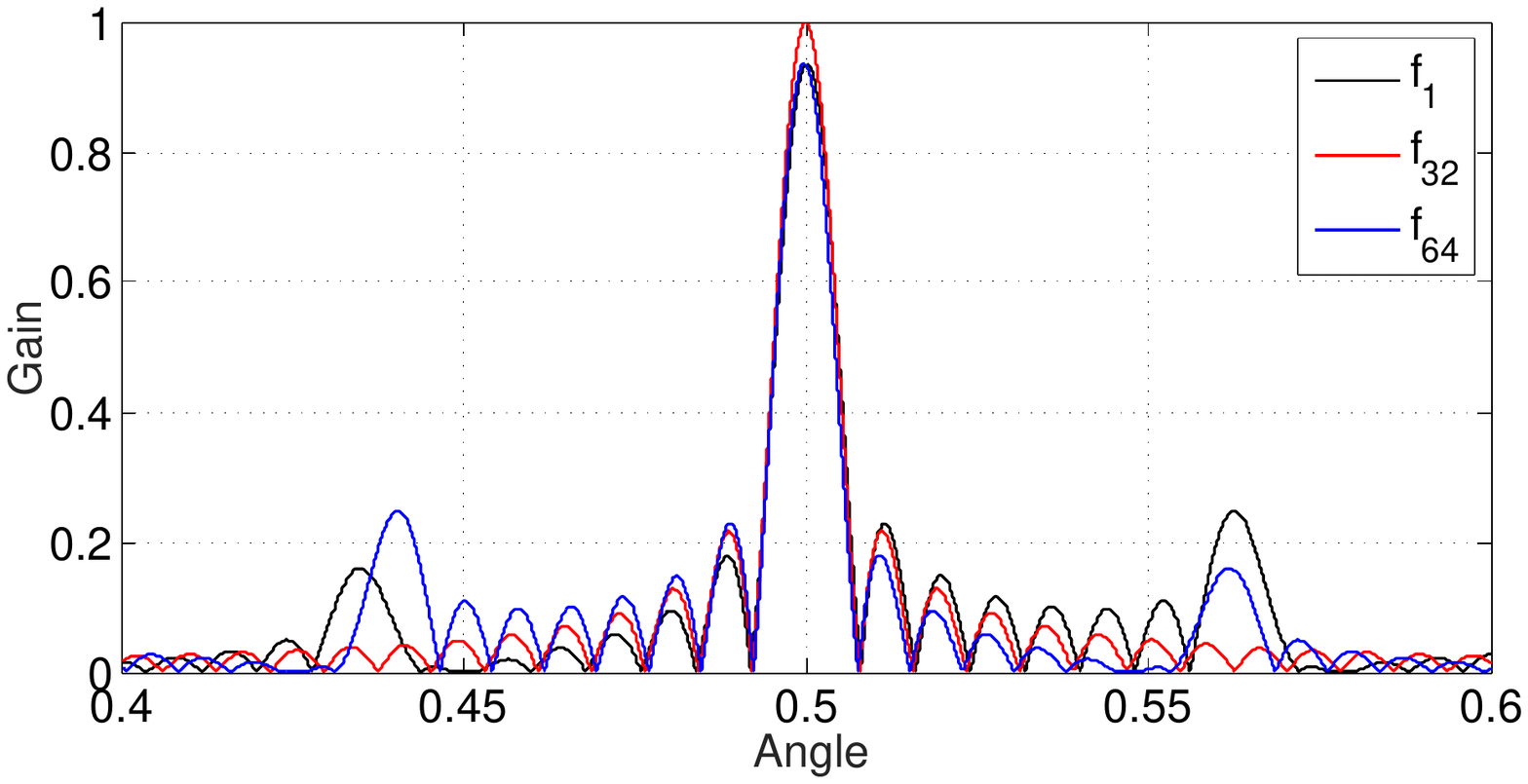}}
	\caption{Beam pattern under (a) conventional subarray antenna structure and (b) TD-based sparse RF chain subarray antenna structure.} 	
	\label{figure5} 
\end{figure}
In this section, we present simulation results to illustrate the system performance under different techniques. First, we show the effect of the IRS deployments on the system performance. As shown in Fig.~\ref{figure3},  we assume that there is a $16\times16=256$ element IRS (i.e., Scheme 1), and then it is transformed into three other schemes, including Scheme 2, Scheme 3, and Scheme 4. Schemes 2 and 3 are used to compare the system performance under different shape deployments, while Scheme 4 is employed to study if more IRSs lead to a better system performance. In addition, we assume that the BS is equipped with a single antenna to serve a single user. The number of subcarriers is 64, while the bandwidth and central subcarrier frequency are $20$ GHz and $200$ GHz, respectively.  Fig.~\ref{figure4} plots the relative subcarrier gain (each subcarrier gain is normalized based on the central subcarrier gain) under the four different IRS deployment schemes. Under the same condition,  one can observe that the relative subcarrier gain is the lowest for the central IRS deployment (Scheme 1), and is the highest for Scheme 4. This means that more IRSs can yield a better system performance. We can also observe from Fig.~\ref{figure4} that Scheme 3 can achieve a higher relative subcarrier gain than Scheme 2. Considering that Schemes 2 and 3 have the same number of IRSs, and  the only difference is the IRS shape, we can conclude that the square IRS deployment is better than the rectangular IRS deployment for the UWB THz IRS communications.

\begin{figure*}
	\centering
	\subfigure[]{
		\label{figure61}	
		\includegraphics[width=5.3cm,height=5cm]{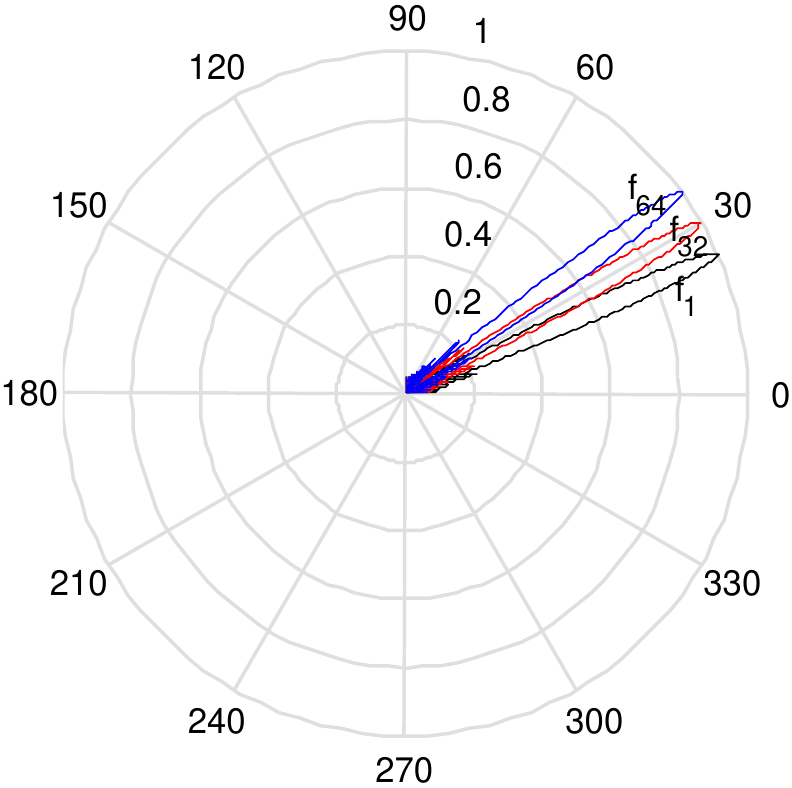}}
	\subfigure[]{
		\label{figure62}	
		\includegraphics[width=5.3cm,height=5cm]{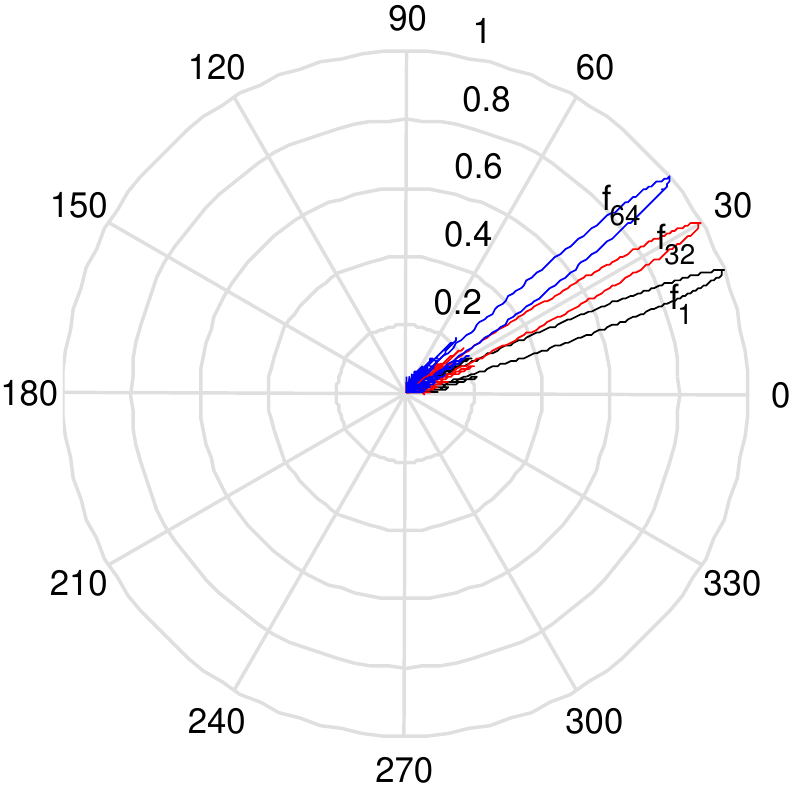}}
	\subfigure[]{
		\label{figure63}	
		\includegraphics[width=5.3cm,height=5cm]{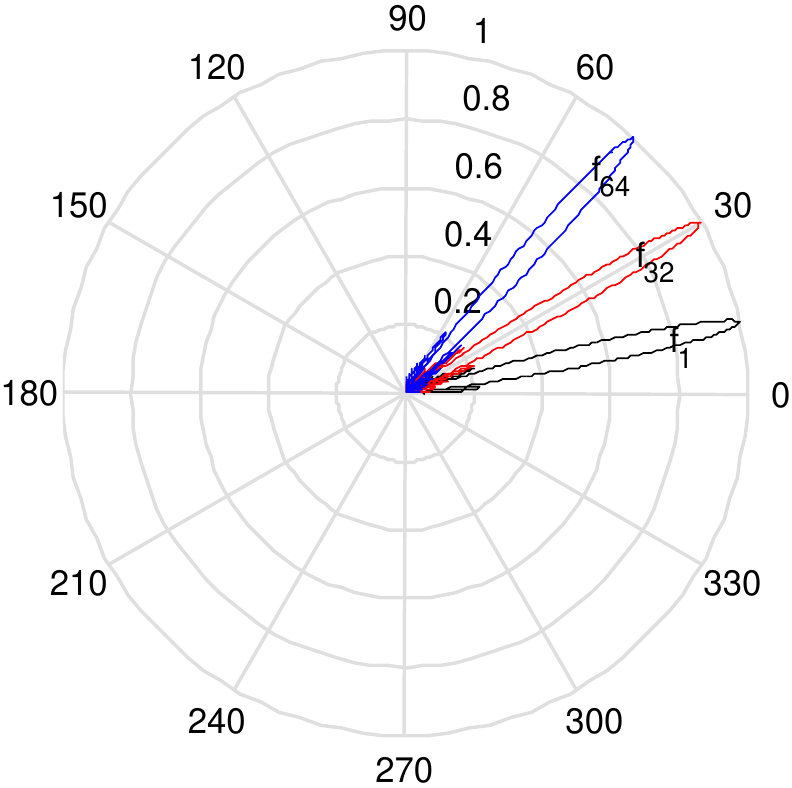}}
	\caption{Beam split model under different TDs, (a) without TDs, (b) 16 TDs, and (c) 32 TDs.}	
	\label{figure6} 
\end{figure*}

Next, we analyze the system performance under the TD-based sparse RF chain antenna structure. Based on the analysis in Section~II, how to overcome the ``beam leakage'' and ``reflecting leakage'' is crucial. It is obvious that the BS should converge the beam for the ``beam leakage'' and broaden the beam  for the ``reflecting leakage''. To realize the above functions, we apply the proposed TD-based sparse RF chain subarray antenna structure (i.e., the third structure in Fig.~\ref{figure2}).  We assume that the BS is only equipped with a single RF chain, and 64 antennas are connected to the RF chain via the multiple TDs and PSs. Here, 4 antennas are grouped to connect to one TD.  Figs.~\ref{figure5}(a) and~\ref{figure5}(b) show the relative gain  with TDs and without TDs, respectively. One can observe that the beam angles of different subcarriers vary from each other when the TDs are not applied as shown in Fig.~\ref{figure5}(a), where they remain the same when the TDs are applied, and thus,  realize the beam coverage. Fig.~\ref{figure6} shows the beam broadening under TDs. Fig.~\ref{figure61} depicts beam split model for the conventional subarray antenna structure, and  Figs.~\ref{figure62} and~\ref{figure63} show the beam broadening model under different numbers of TDs, where 16 TDs and 32 TDs are used, respectively. It is clear that more TDs can achieve a better performance in beam broadening. However, the TD elements are usually of high power consumption and hardware complexity. 

\section{Open Issues and Research Opportunities}
We highlight several open issues for the UWB THz IRS communications, and discuss the corresponding research directions and opportunities  in this section.  
\subsection{AI/ML-Empowered Techniques}
{Due to ultra-high frequency and large array antennas at the BS and IRS, how to obtain CSI is a critical and challenging problem. Furthermore, joint beamforming design is also non-trivial due to the two-stage beam split. To address the above challenges, a potential solution is to apply AI/ML techniques. For example, by exploiting the channel sparsity, advanced neural networks can be designed to obtain the CSI with lower complexity. Consequently,  intelligent beamforming and antenna selection can be applied to solve the complicated beamforming design problem. However, how to utilize the AI/ML techniques for THz IRS communications more effectively should be extensively investigated in the future.} 
\subsection{Selection of the TDs}
As shown in Fig.~\ref{figure2}, several TD-based sparse RF chain antenna structures are designed. It is obvious that the TD is very important for overcoming the first stage beam split effect, and more TDs can result in higher performance. However, the large number of TDs will lead to high power consumption and hardware complexity, and thus,  is difficult and inapplicable in practical hardware design. Therefore, we need to select the appropriate number of  TDs to obtain a tradeoff between system performance and hardware complexity.
\subsection{Discrete Phase Shift Design at the BS and IRS}
For UWB THz IRS communications, when the BS uses a sparse RF chain antenna structure, the analog beamforming needs to be designed by adjusting the phase shifts, and the IRS reflects the received signals to form the directional beams by adjusting the phase shifts. Generally, continuously tuning the phase shifts at the BS and IRS can yield near optimal performance, and this is convenient to analyze. However, it is difficult and costly to implement in practice because of the complex hardware design. For example, when 16 levels of phase shift are required, 4 PIN diodes are needed for each element. With a large number of elements at the BS or IRS, it is very challenging to pack so many diodes within the limited physical space,  and the controlling is also an issue. Accordingly, two stage discrete phase shifts at the BS and IRS should be jointly optimized, which is a new challenge.
\subsection{Semantic Communications}
{{Current communication technologies are mainly based on the Shannon physical capacity, and they are limited by the advanced encoding and modulations techniques. For future 6G wireless networks, semantic communications is a promising technique, and its core is to directly transmit ``the meaning'' of the information without encoding/decoding or modulation, which improves the transmission efficiency. However, how to apply this advanced technique in UWB THz IRS communications is a fruitful future research direction.}}    
\subsection{Sensing and Localization} 
{{In addition to provide the ultra-high data transmission rate, the  THz frequency band is also attractive for high resolution sensing and localization due to its transmission characteristic, such as the fast tracking and millimeter-resolution wireless positioning. Furthermore, the IRS is also helpful for  wireless sensing and localization. Therefore,  how to integrate the communication, sensing and localization in THz IRS communication is a challenging and meaningful research direction.}}
\subsection{Simultaneous Transmitting and Reflecting IRS Communications}
To improve the efficiency of IRS communications, simultaneous transmitting and reflecting IRS (STAR-IRS) has been developed, which means that the IRS can simultaneously transmit and reflect the incident signals, thus improving coverage and transmission efficiency.  However, for the UWB THz IRS communications, several issued should be explored, such as how to analyze the beam split for the transmit beamforming formed at the IRS, its effects on performance, and how to jointly optimize the transmission/reflection coefficients, and the IRS deployment.     
\subsection{Joint Active and Passive IRS Communications}
Although a passive IRS can decrease the energy consumption and hardware complexity, the ``double fading'' effect yields a small achievable capacity gain. To solve this problem, active IRSs are considered in several existing works, where the signals can be amplified  to break the ``double fading'' effect, thus improving the transmission efficiency. However, fully active IRS may cause large power consumption. Accordingly, there exists the tradeoff between the transmission efficiency and energy consumption and how to analyze joint active and passive IRS are promising directions. Furthermore, the beam split effect can be reduced by active IRS, which should also be explored. 
\section{Conclusion}
{In this article, we studied UWB THz IRS communications, and introduced several potential  application scenarios. We presented major challenges and key techniques, where we carefully analyzed the two stage beam split and the effect of the IRS deployment on the system performance. Furthermore, we proposed a novel antenna structure with low hardware complexity and energy consumption, and validated its effectiveness via simulation results.   Finally, we highlighted several key research directions and promising opportunities for future UWB THz IRS communications.}

\end{document}